\documentclass[aps, preprint, onecolumn,secnumarabic,nobalancelastpage,nofootinbib,superscriptaddress]{revtex4-1}
\usepackage{graphicx}
\usepackage{amsmath,amssymb}
\usepackage[utf8]{inputenc}
\usepackage{natbib}
\usepackage{hyperref}
\usepackage{color}
\hypersetup{colorlinks=true,citecolor={blue},linkcolor={blue},urlcolor={blue}}
\usepackage{units}
\usepackage[english]{babel}
\usepackage{upgreek}
\usepackage[normalem]{ulem}
\usepackage{soul,xcolor}

\begin{document}
\title{Quantum fluids of light in all-optical scatterer lattices}

\date{\today}
\author{S. Alyatkin}
\affiliation{Skolkovo Institute of Science and Technology, Moscow, Bolshoy Boulevard 30, bld. 1, 121205, Russia}
\author{H. Sigurdsson}
\affiliation{Skolkovo Institute of Science and Technology, Moscow, Bolshoy Boulevard 30, bld. 1, 121205, Russia}
\author{A. Askitopoulos}
\affiliation{Skolkovo Institute of Science and Technology, Moscow, Bolshoy Boulevard 30, bld. 1, 121205, Russia}
\author{J. D. T\"opfer}
\affiliation{Skolkovo Institute of Science and Technology, Moscow, Bolshoy Boulevard 30, bld. 1, 121205, Russia}
\author{P. G. Lagoudakis}
\email{P.Lagoudakis@skoltech.ru}
\affiliation{Skolkovo Institute of Science and Technology, Moscow, Bolshoy Boulevard 30, bld. 1, 121205, Russia}

\begin{abstract}
One of the recently established paradigms in condensed matter physics is examining a system's behaviour in artificially constructed potentials, giving insight into physical phenomena of quantum fluids in hard-to-reach settings. A prominent example is the matter-wave scatterer lattice, also known as the barrier lattice or repulsive Dirac comb. There, high energy matter waves undergo transmission and reflection through narrow width barriers leading to stringent phase matching conditions with subsequent lattice band formation. It is one of the most well taught system in quantum mechanics but its realisation for macroscopic matter-wave fluids has remained elusive, in contrast to evanescently coupled lattice sites or waveguides. Here, we implement and study a system of exciton-polariton condensates in a non-Hermitian Lieb lattice of scatterer potentials by optically injecting incoherent exciton clouds which both emit, and interact with traveling polariton waves. By fine tuning the lattice parameters, we reveal a nonequilibrium phase transition between two distinct regimes of polariton condensation: a scatterer lattice of gain guided polaritons condensing on the lattice potential maxima, and trapped polaritons condensing in the lattice potential minima. The transition is characterised by multimodal condensation due to gain competition between the two regimes. Energy tomography on the polariton emission enables us to measure the intricate band structure of the optically induced lattices. Our results pave the way towards unexplored physics of non-Hermitian fluids in non-stationary mixtures of confined and freely expanding waves. 
\end{abstract}


\maketitle
\section*{Introduction}
Artificial lattices are highly attractive for obtaining insight into properties of crystal structures in the solid state, and for creating patterned structures not found in nature. They can be used to investigate spin frustration in triangular geometries~\cite{Skjaervo_NatRev2020}, massless Dirac fermions in honeycomb structures~\cite{Wang_NatNano2018}, and strongly correlated states in dispersionless flat bands~\cite{Polini_NatureNano2013}. Conventionally, the physical properties of lattices are investigated through an appropriate choice of confined single-particle states (Wannier functions), such as electrons bound to their atoms, ultracold atoms in optical traps~\cite{Schafer_NatRev2020}, or index-guided electromagnetic waves in photonic crystals~\cite{Joannopoulos_photonic_crystals}. The evanescent transfer of energy between lattices sites is then captured within the tight-binding approximation to explain crystal band formation, the centerpiece of solid-state physics~\cite{ashcroft_solid}.
 
However, in contrast to lattices of tightly confined states there exists the inverse case of coherent matter-wave scattering in the diffractive regime of lattices made up of repulsive potentials much smaller in size than the lattice spacing. There, waves with wavelength smaller than the lattice constant experience strong reflection and diffraction from the lattice (e.g., Bragg's law). Since the early work of Kronig and Penney~\cite{Kronig_PRSL1931}, an ordered arrangement of spherically symmetric scatterers is known to give rise to bands and bandgaps. The study of elastic scattering of incoming and outgoing states on a static object is at the heart of mechanical, electromagnetic-, Schr\"{o}dinger-, and sound-wave scattering theory~\cite{Newton_ScatteringTheory}. However, the realisation of lattices of scatterers acting on matter waves with coherence length exceeding the scatterer spacing is much harder than the implementation of lattices of confined (evanescently coupled) wavefunctions, and the connection between the two regimes remains mostly unexplored.
\begin{figure}[b]
	\includegraphics[width = \textwidth]{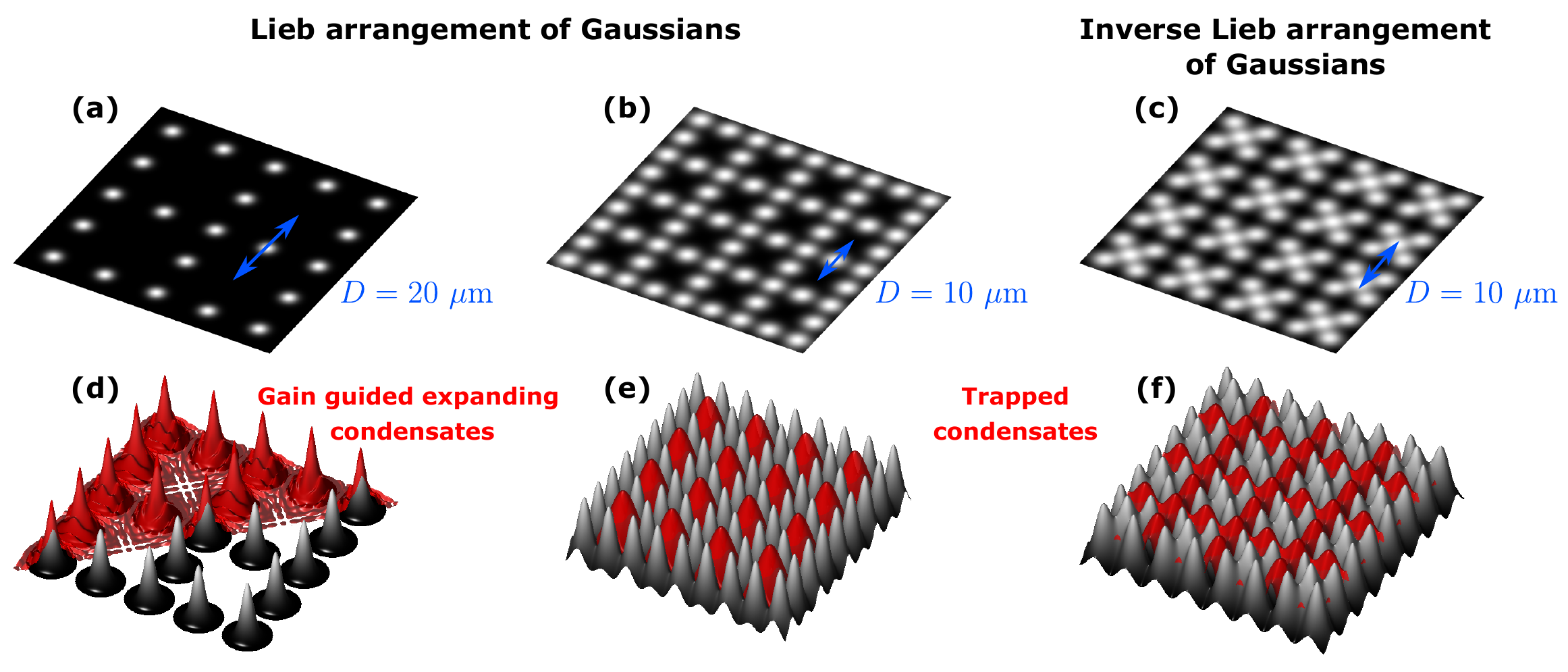}
	\caption{\textbf{Schematic of the optical excitation pattern and resulting polariton condensates.} Excitation intensity profile composed of Gaussian pump spots arranged in (a,b) a Lieb pattern for two different lattice constants $D$ and (c) an inverse Lieb pattern with the potential minima (dark areas) forming a conventional Lieb lattice. (d-f) Corresponding black-white surface plots of the pump induced potential landscapes with the polariton condensate density overlaid as red envelope. In (d) polariton condensation occurs on the potential maxima (gain guided condensates) resulting in highly energetic (ballistic) condensate waves whereas in (e,f) condensation takes place in the potential minima between pump spots.}
	\label{fig1}
\end{figure}

Semiconductor microcavities in the strong coupling regime are especially appealing for engineering artificial lattices as they host matter-wave modes known as exciton-polaritons (from here on {\it polaritons})~\cite{kavokin_microcavities_2007}. These modes possess large nonlinearities, picosecond scale response times, and permit easy optical write-in and read-out of information. Polaritons can undergo power-driven nonequilibrium Bose-Einstein condensation, making them favourable candidates to study low threshold room-temperature lasing~\cite{Christopoulos_PRL2007}, optical many-body hydrodynamic phenomena~\cite{Stepanov_NatComm2019, Suarez_LSA2020}, topological phases~\cite{klembt_exciton-polariton_2018, Gianfrate_Nature2020}, and implementation of optical information processing~\cite{Zasedatelev_NatPho2019}. Moreover, strong interparticle interactions result in repulsion of condensate polaritons from a background of uncondensed particles (i.e., photoexcited exciton reservoir), co-localised with the pumped area~\cite{Wertz_NatPhys2010, askitopoulos_polariton_2013}. This enables all-optical design of non-Hermitian (i.e., optical gain and blueshift) potential landscapes for polaritons~\cite{tosi_sculpting_2012, alyatkin_PRL, Pickup_NatComm2020}, in analogy to dipole optical traps for cold atoms~\cite{Bloch_NatPhys2005} or photorefractive crystals~\cite{Neshev_PRL2004}. Indeed, with non-Hermitian potentials the wavefunction norm is no longer conserved and, in conjunction with stabilising condensate nonlinearities, synchronisation can spontaneously appear amplifying the matter-wave similarly to phase-locked laser arrays~\cite{Glova_QuaElec2003, Mawst_IEEE2003}.

Here, we demonstrate a non-resonantly optically imprinted repulsive (scatterer) lattice (see Figs.~\ref{fig1}a,d) wherein scattered high energy polariton waves, emitted from the pump areas, result in robust interference patterns due to their ability to dynamically adjust their phase in order to condense into the highest gain Bloch state. Our scatterer lattice is chosen to have the edge-centred square (Lieb) arrangement, a configuration not found usually in nature, which offers comparison against the conventional tight-binding Lieb lattice (see Figs.~\ref{fig1}c,f) which we also all-optically engineer. We perform full momentum-energy space tomography to unveil the engineered lattice band structures and their reshaping by altering the lattice parameters. We observe a gradual nonequilibrium phase transition from the scatterer lattice of ballistically expanding polariton waves (Figs.~\ref{fig1}a,d) to a tight-binding lattice of trapped condensates (Figs.~\ref{fig1}b,e) bridged by an unstable regime of multimodal condensation due to gain competition. Moreover, underscoring the flexibility of our optical approach, we provide direct observation of dispersionless P-flatband condensation achieved by using an excitation profile forming an ``inverse'' Lieb lattice (Figs.~\ref{fig1}c,f) in the same spirit as the vacancy lattice created in electronic systems~\cite{Drost_NatPhys2017, Slot_NatPhys2017} or optical lattices of cold atoms~\cite{Shen_PRB2010, Gross_Science2017}. We point out that the majority of our findings are not strongly dependent on the choice of Lieb lattice arrangement and can be extended to other types such as square, honeycomb, and triangle lattices.

\section*{Results}
 A strain compensated 2$\lambda$ GaAs-based planar microcavity with embedded three pairs of In$_{0.08}$Ga$_{0.92}$As quantum wells \cite{cilibrizzi_polariton_2014} and an exciton-photon detuning of $-4\;\mathrm{meV}$ is held at $\approx 4\;\mathrm{K}$ in a closed-cycle helium cryostat. The non-resonant excitation (single mode laser tuned at 1.5578 eV) is amplitude modulated at a frequency of 5 kHz (duty cycle 1\%) with an acousto-optical modulator to avoid sample heating. A desired pump profile is shaped by a computer-controlled reflective phase only spatial light modulator~\cite{toepfer_coherence2020} and projected onto the sample through a microscope objective (NA=0.42). We collect the real space polariton photoluminescence (PL), directly proportional to the condensate density, in transmission geometry and spectrally filter it from the excitation laser. We denote the horizontal and vertical momentum space coordinates as $\mathbf{k} = (k_x,k_y)$, corresponding to the spatial frequencies along horizontal ($x$-axis) and vertical ($y$-axis) real space coordinates, respectively.

As mentioned above, the flexibility in tuning the lattice properties through optical excitation structuring makes a system of microcavity polaritons appealing to access various lattice physics in a recyclable setting. The height of the potential landscape is determined by the excitation intensity and, in this study, constructed by an arrangement of Gaussian shaped pump spots. On one hand, in the scatterer lattice where the pump spots have large separation distances with multiples of the polariton wavelength, polaritons condense on the maxima of the pumped potentials (see Figs.~\ref{fig1}a,d) characterised by gain guided bright centres and ballistically expanding envelopes~\cite{Pickup_NatComm2020, toepfer_2020, alyatkin_PRL, toepfer_coherence2020}. On the other hand, for closely spaced pump spots, they can condense in the minima of the potential landscape (see Figs.~\ref{fig1}b,e and~\ref{fig1}c,f) becoming optically trapped~\cite{Cristofolini_PRL2013, ohadi_synchronization_2018, pieczarka_arxiv2021} partly because of their strong interactions that help them relax in energy. In the former case, the system shares analogies with with antiguided laser arrays~\cite{Mawst_IEEE2003}, whereas in the latter case with two-dimensional (2D) electron transport~\cite{Wang_NatNano2018}.
\begin{figure*}[t!]
	\includegraphics[width=\linewidth]{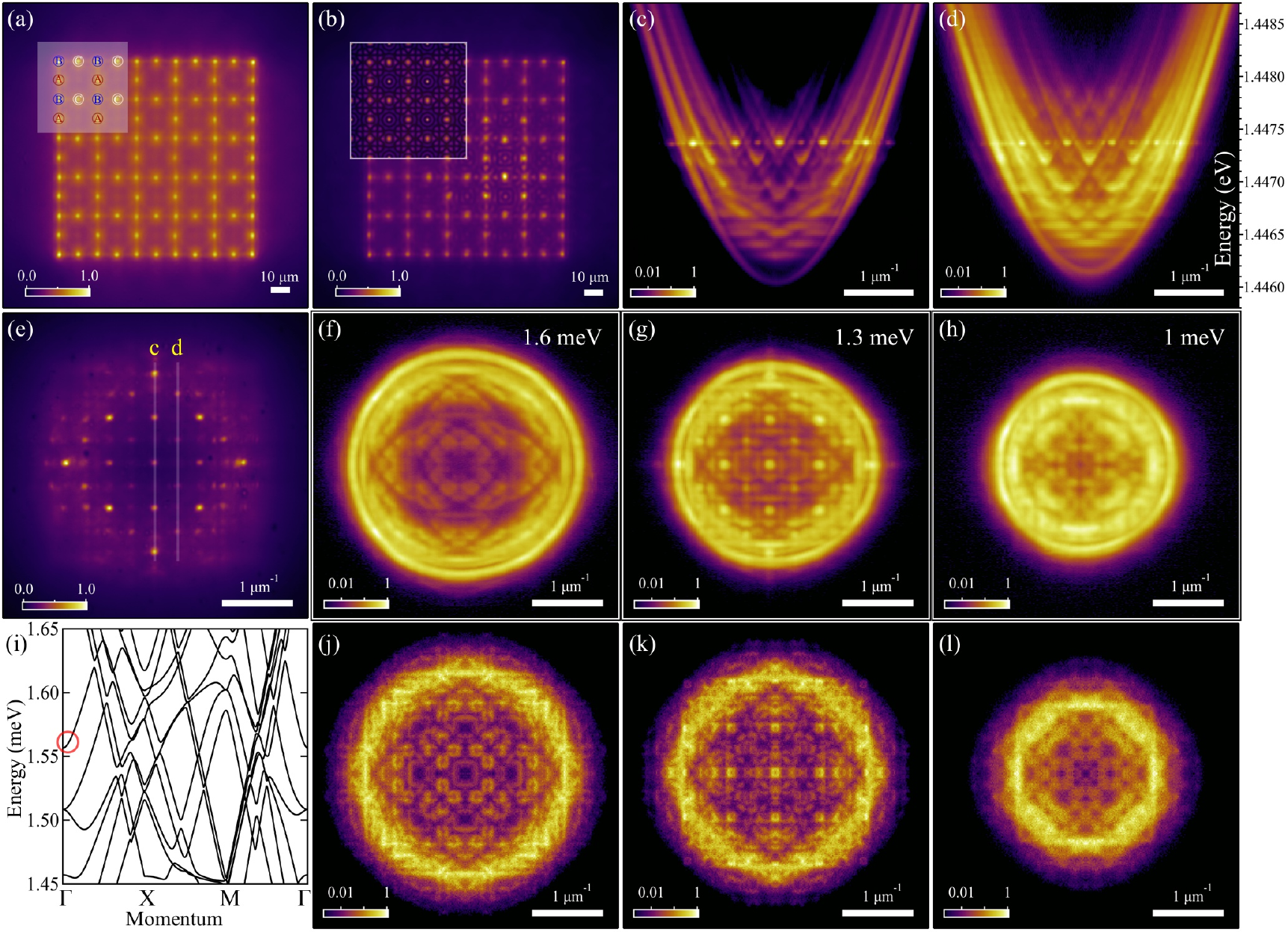}
	\caption{(a) Measured PL from a scatterer Lieb lattice of polariton condensates with a lattice constant set to  $D = 20.3$ $\upmu$m excited at $P= P_{\mathrm{thr}}$ and (b) at  $P=1.2 P_{\mathrm{thr}}$. Inset in (b) shows the calculated Bloch state density for comparison. The inset in (a) shows sublattices denoted with letters A, B, C forming the lattice. (c,d) Energy-resolved momentum space polariton PL at $P=1.2 P_{\mathrm{thr}}$  for $k_x$=0 and $k_x= 2\pi/D$ values respectively, marked on (e) with vertical lines and yellow letters ``c'' and ``d''. (e) Energy integrated momentum space polariton PL at  $P=1.2 P_{\mathrm{thr}}$, and (f,g,h) energy-resolved ``slices'' of momentum space at energies 1.6 meV, 1.3 meV and 1 meV with respect to the bottom of the lower polariton branch (dispersion). (i) Calculated lattice bands with the red circle corresponding to the Bloch state in the inset in (b). (j,k,l) Numerically calculated ``slices'' of polariton momentum space PL from Monte-Carlo simulations on the 2D dissipative Schr\"{o}dinger equation.}
	\label{fig2}
\end{figure*}

In Fig.~\ref{fig2}a we show the real space PL of an optically pumped scatterer lattice with 96 polariton condensates at threshold power ($P=P_\text{thr}$) arranged into a Lieb geometry (corresponding to Figs.~\ref{fig1}a,d). Each condensate is strongly gain guided with a bright centre co-localised with its respective pump spot. The size of the Gaussian pump spots sustaining the condensates is $\approx 2$ $\upmu$m full width at half maximum and the lattice constant is set to $D = 20.3\ \upmu$m. As schematically shown in the inset of Fig.~\ref{fig2}a, the Lieb lattice is composed of three square sublattices denoted with the letters A, B, and C. Being close to threshold, the spatial coherence of each condensate does not extend outside its respective pump spot~\cite{Roumpos_PNAS2012} making them isolated (uncoupled) objects. Figures~\ref{fig2}b and~\ref{fig2}e show the real and momentum space polariton PL above threshold ($P = 1.2 P_\text{thr}$), while Figs.~\ref{fig2}c and~\ref{fig2}d show energy-momentum space PL along $k_x = 0$ and $k_x = 2\pi/D$, respectively. By driving the system above threshold we increase both the coherence and particle outflow from each condensate resulting in stronger coupling between the radiating condensates and the appearance of interference fringes (signature of synchronisation) as well as distinct Bragg diffraction peaks in momentum space. Specifically, we observe the condensates populating an excited Bloch state characterised by a constructive interference peak in the centre of each lattice cell, also visible from the calculated Bloch state shown in the inset in Fig.~\ref{fig2}b. By scanning the Fourier space PL (collected in reflection geometry) image across a slit of a spectrometer with a motorised mount,  we reconstruct energy-resolved ``slices'' (i.e., isoenergy planes) of momentum space PL at three specific energies, shown in Figs.~\ref{fig2}f-h. The applied energy tomography reveals the full picture of complex band formation (see Supplementary Movie 1). We also calculate the band structure using two different modeling methods: Bloch's theorem, and Monte-Carlo sampling of the 2D dissipative polariton Schr\"{o}dinger equation [see Secs.~S2 and~S3 in the Supplemental Material (SM) for details]. The Bloch analysis, as shown in Fig.~\ref{fig2}i, reveals a zoo of overlapping bands of distinct shapes, yet the polaritons, being so interactive, are still easily able to relax into a definitive Bloch state corresponding to the optimum gain (marked with a red circle and plotted in Fig.~\ref{fig2}b inset), in a similar spirit to coupled laser systems. The numerical Monte-Carlo sampling of the dominant Fourier components of the scatterer lattice shown in Figs.~\ref{fig2}j-l gives good agreement with the experiment. We note that the illuminated, clearly formed, bands in Figs.~\ref{fig2}c and ~\ref{fig2}d show that polaritons in the repulsive scatterer Lieb lattice indeed experience crystal scattering within their coherence time. We also observed band-structure formation for the square scatterer lattice of 5$\times$5 polariton condensates with a similar lattice constant (see Sec.~S1 in SM).

\begin{figure*}[t]
	\includegraphics[width=0.95\linewidth]{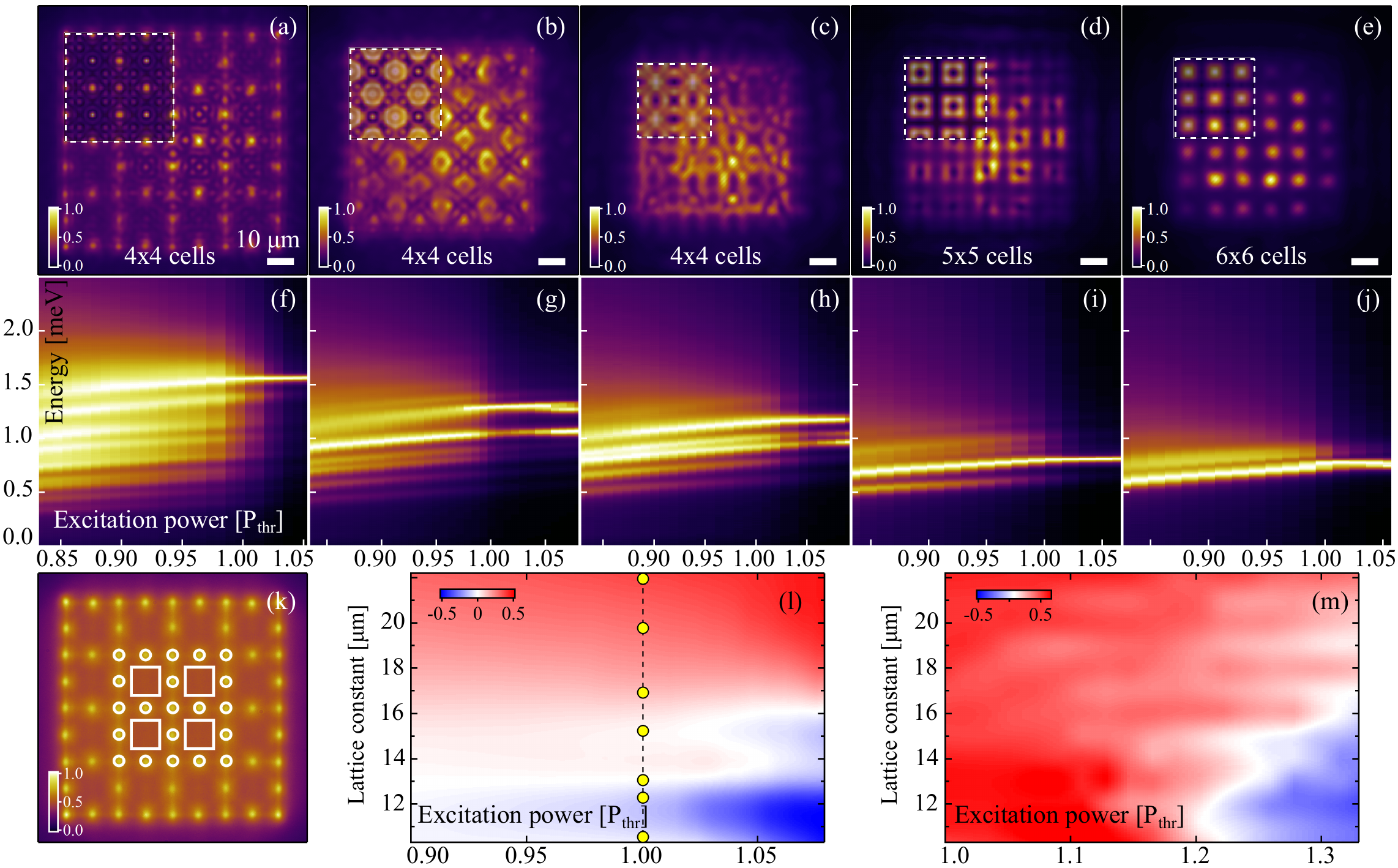}
	\vspace{-10 pt}
	\caption{\textbf{Optical lattices of polariton condensates pumped with Lieb geometry for different lattice constants $D=$ 20.0, 16.9, 15.2, 12.3, 10.3 $\upmu$m}. (a-e) show corresponding real space polariton PL above condensation threshold and (f-j) corresponding spectra as functions of pump power (energy is scaled with respect to the bottom of lower polariton branch). Semi-transparent insets in (a-c,e) marked with white dashed squares show results of time-averaged numerical simulations of the condensate dynamics using the generalised Gross-Pitaevskii equation. Inset in (d) is calculated using Bloch's theorem. White circles and squares overlaid with real space PL below threshold for $D = 20\ \upmu$m in (k) denote spatial integration areas on top of and outside the pump spots used to extract the contrast between gain guided ($I_\text{spots}$) and trapped ($I_\text{trapped}$) polariton PL. This contrast is plotted as normalised heatmaps $S = (I_\text{spots}-I_\text{trapped})/(I_\text{spots}+I_\text{trapped})$ in (l,m) as a function of lattice constant and pump power from experiment and simulations, respectively, revealing a gradual transition between the two regimes. Scale bar in (a) applies also to (b-e) and (k). Yellow circles in (l) denote the lattice constants realized experimentally.}
	\label{fig3}
	\vspace{-10 pt}
\end{figure*}

Next, we decrease the lattice constant from $D=20$ $\upmu$m to $D = 10.3$ $\upmu$m and characterise the change in the polariton system behaviour. Corresponding images with experimentally measured real space polariton PL above condensation threshold are depicted in Figs.~\ref{fig3}a-e. As shown in Fig.~\ref{fig3}a, for $D= 20$ $\upmu$m, most of the PL intensity comes from the pumped areas with clear interference fringes in time-averaged measurements as a consequence of the scattered polariton waves leading to robust synchronisation between the condensates~\cite{toepfer_2020, toepfer_coherence2020}. Multiple weakly populated energy branches collapse into one dominant mode as the pump power exceeds threshold value, as clearly seen in Fig.~\ref{fig3}f. Decreasing the lattice constant to $D = 16.9$ $\upmu$m results in substantial growth of PL intensity inside each lattice cell (i.e., where a Gaussian pump spot is absent) with simultaneous decreased PL at the pumps positions (see Fig.~\ref{fig3}b). Driving the system above threshold leads to dual mode condensation as confirmed by the measured spectrum power scan shown in Fig.~\ref{fig3}g. Decreasing the lattice constant to $D = 15.2$ $\upmu$m again dramatically modifies the polariton PL pattern. As shown in Fig.~\ref{fig3}c, polaritons are repelled even stronger outside the pumped areas leading to complex PL distribution in real space. The condensate is here fractured into multiple energy modes above threshold (see Fig.~\ref{fig3}h), with similar gain. Finally, decreasing to even smaller lattice constants of $D = 12.3$ $\upmu$m and $D = 10.3$ $\upmu$m results in formation of trapped condensates~\cite{ohadi_synchronization_2018} as shown in Figs.~\ref{fig3}d and~\ref{fig3}e, characterised by a dominant single energy mode above threshold (see Figs.~\ref{fig3}i and~\ref{fig3}j). This regime is schematically depicted in Figs.~\ref{fig1}b and~\ref{fig1}e.

The most intriguing physics lies in the intermediate regime where multimodal condensation takes place between the regimes of gain guided ballistically coupled condensates (Fig.~\ref{fig3}a) and trapped condensates (Figs.~\ref{fig3}d and~\ref{fig3}e). In order to distinguish these regimes and identify the transition between them we introduce a contrast parameter $S= (I_\text{spots} - I_\text{trapped})/(I_\text{spots} + I_\text{trapped})$ describing the normalised difference between the average polariton PL at the pumping spots $I_\text{spots}$ (white circles in Fig.~\ref{fig3}k) and outside the spots $I_\text{trapped}$ (white squares in Fig.~\ref{fig3}k). Here, ``average'' refers to area-integrated PL divided by the integration area. The side length of the white squares is chosen as $L = 0.54D$ (the precise value does not affect the findings) whereas the diameter $d = 3.6\;\mathrm{\upmu m}$ of the white circles is fixed for all lattice constants. 

In case of well separated pump spots (i.e., $D>17$ $\upmu$m) we observe values $S>0$ in Fig.~\ref{fig3}l across all pump powers corresponding to high-energy gain guided polaritons that scatter across the lattice when emitted from their pump spots. In the opposite case of small separation distances between the pumps (i.e., $D<12.5$ $\upmu$m) we observe $S < 0$ corresponding to polaritons becoming trapped in the lattice potential minima with dominant PL intensity coming from the white squares in Fig.~\ref{fig3}k. Here, high energy scattering of polaritons across multiple lattice cells is reduced, replaced with evanescent transfer of energy (i.e., tunneling). Therefore, at some lattice constant (where $S \approx 0$) the polariton system undergoes a nonequilibrium phase transition between these different condensation regimes. During this transition, competing lattice modes fight over the gain which leads to a ``frustrated`` condensation pattern, like shown in Figs.~\ref{fig3}b and~\ref{fig3}c. This transition is unique to polariton systems since it is the excitonic part of polaritons which allows them to interact with each other and the uncondensed exciton reservoir, which facilitates energy relaxation~\cite{Wertz_NatPhys2010}. In contrast, purely photonic systems would generally lase in their pumped gain region. The gradual onset of trapped polaritons as the lattice constant reduces and pump power increases (i.e., the background uncondensed reservoir increases) can therefore be attributed to modes between the pump spots moving into resonance with the lattice gain bandwidth due to enhanced energy relaxation of polaritons~\cite{Wouters_PRB2010}. We point out that our all-optical experiment and extraction of $S$ for different lattice constants and pump powers is not possible to replicate with lithographically written photonic periodic structures where the pumped gain region would correspond to the lattice potential minima~\cite{klembt_exciton-polariton_2018}. In our experiment, the pump gain follows the lattice potential maxima. We qualitatively reproduce our findings through numerical mean-field simulations using the generalised Gross-Pitaevskii equation~\cite{Wouters_PRL2007} including an energy relaxation mechanism~\cite{Wouters_PRB2010} (see Sec.~S4 in SM for details), shown in the insets of Fig.~\ref{fig3}a-c,e and Fig.~\ref{fig3}m. The state in the inset in Fig.~\ref{fig3}d was calculated using Bloch's theorem and did not appear in stable form in Gross-Pitaevskii simulations for our chosen (fixed) set of simulation parameters. Some discrepancy can be observed between simulation and experiment in Figs.~\ref{fig3}l and~\ref{fig3}m at low powers and small lattice constants where the simulation overestimates the gain guided polaritons. This discrepancy could be reduced by applying a stochastic treatment (e.g., Wiener noise) to the condensate equations of motion which would smear out the simulated condensate PL at low powers close to threshold, or by including exciton diffusion in the model.
\begin{figure*}[t!]
	\includegraphics[width=0.85\linewidth]{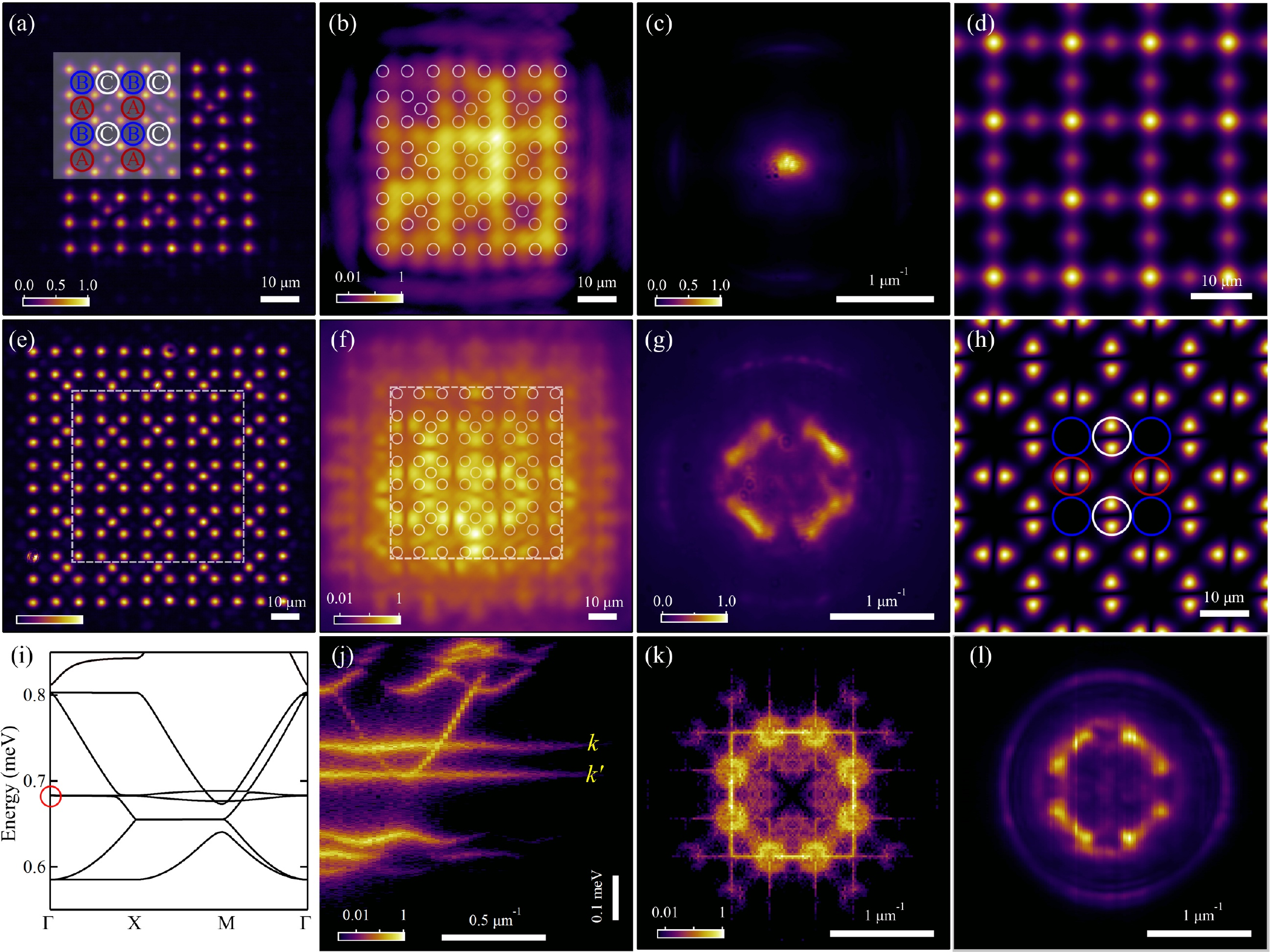}
	\caption{\textbf{Optical inverse Lieb lattice of Gaussians with flatband condensation.} (a,e) Spatial pumping profile of Gaussian spots arranged in the inverse Lieb geometry with lattice constants $D = 13$ $\upmu$m and $16$ $\upmu$m, respectively. In (a) the cube-centred spots are 40\% weaker in intensity compared to the rest of the pump spots. In (e) all spots are equally intense. (b,f) Real and (c,g) momentum space polariton PL above condensation threshold corresponding to populated S-band ground state and P-flatband state, respectively. White circles in (b) and (f) denote the pumps positions. (d,h) Corresponding calculated real space wavefunction densities intensities from Bloch's theorem. Coloured circles in (a) and (h) denote the Lieb lattice unit cells. (i) Calculated P-bands from Bloch's theorem where the red circle marks the flatbands. (j) Calculated dispersion cross-section at $k_x= 2\pi/D$ from Monte-Carlo Schrödinger numerics. Both (i) and (j) use the pump profile shown in (e). (k,l) Calculated and measured PL from the isoenergy plane in momentum space corresponding to the P-flatband marked with the yellow letter \textit{k} in (j). The band marked with the yellow letter \textit{k'} is the lowest energy P-band which only appears flat between the $\Gamma$-$X$ points in the Brillouin zone.}
	\label{fig4}
\end{figure*}

To further demonstrate the versatility of our all-optical approach we move away from the scatterer Lieb lattice and design now the conventional Lieb lattice (see Figs.~\ref{fig1}c and~\ref{fig1}f) which has been studied vigorously in various systems of electrons~\cite{Drost_NatPhys2017, Slot_NatPhys2017}, photons~\cite{Guzman_NJP2014, Mukherjee_PRL2015, Vicencio_PRL2015}, cold atoms~\cite{Shen_PRB2010, Taie_SciAdv2015}, plasmon polaritons~\cite{Kajiwara_PRB2016}, and exciton-polaritons~\cite{Klembt_APL2017, Whittakes_PRL2018, Goblot_PRL2019, Harder_PRB2020, Scafirimuto_CommPhys2021}. It hosts an excess of phenomena including topological phases, dispersionless flat bands, and Dirac points making it a popular testbed in solid-state physics. In Fig.~\ref{fig4} we realise the conventional Lieb lattice by arranging the Gaussian excitation spots in an inverse Lieb lattice (Figs.~\ref{fig4}a and~\ref{fig4}e). Just as in Fig.~\ref{fig3}d and~\ref{fig3}e, here the polariton condensates are designed to be trapped in the lattice potential minima, interacting via tunnelling. Figures~\ref{fig4}b and~\ref{fig4}c show the real- and momentum space polariton PL above threshold, respectively, in a $D = 13$ $\upmu$m lattice revealing condensation into the S-band ground state in the $\Gamma$ point at $k=0$. By adjusting the lattice parameters, the condensate can be forced to populate a different Bloch state. Figures~\ref{fig4}e-g show the same measurements but for $D = 16$ $\upmu$m where we now observe condensation into P-orbitals at sublattices A and C, whereas at sublattice B destructive interference occurs. This state belongs to a dispersionless P-flatband~\cite{Klembt_APL2017, Whittakes_PRL2018} and constitutes the first evidence of polariton condensation into optically (non-Hermitian) generated flatband states. Figures~\ref{fig4}d and~\ref{fig4}h show corresponding calculated lattice states using Bloch's theorem. The state plotted in Fig.~\ref{fig4}h corresponds to the band denoted by the red circle in Fig.~\ref{fig4}i where we show the calculated P-bands along the reduced Brillouin zone edge. We additionally calculate the dispersion of the lattice from Monte-Carlo sampling of the dissipative Schrödinger equation and plot the results in Fig.~\ref{fig4}j along the $k_x = 2\pi/D$ direction. The results reveal bright bands where polaritons decay slowest with the P-flatband clearly visible (marked with yellow {\it k}) and the lowest energy P-band (marked with yellow {\it k'}) which only appears flat along the $\Gamma$-$X$ direction.

Experimentally implemented energy tomography measurements allow us to extract PL belonging to isoenergy planes in reciprocal space which further verifies that the condensate is populating the P-flatband. Figures~\ref{fig4}k and~\ref{fig4}l show the calculated and measured PL from the P-flatband isoenergy plane, respectively, which is marked with the yellow colored letter \textit{k} in Fig.~\ref{fig4}j. Indeed, the similarity between Fig.~\ref{fig4}l and the energy-integrated PL in Fig.~\ref{fig4}g confirms that condensate is dominantly populating a single energy state with similar intensity pattern obtained from calculation (see Sec.~S1 in SM for more details). It should be noted that the calculated PL (Fig.~\ref{fig4}k) shows some finer details in momentum space which would average out in time-integrated measurements due to non-ideal effects such as noise, cavity disorder, and pump fluctuations. However, the qualitative structure is the same as in experiment with most of the PL belonging to the $M$ symmetry points. The fact that the PL is strongest at the $M$ points can be understood from the slight curvature in the P-flatband around these points (see Fig.~\ref{fig4}i) which facilitates the relaxation of polaritons into the band minima.

We now address the question on why condensation occurs into the P-flatband as shown in Figs.~\ref{fig4}f and~\ref{fig4}g. As mentioned in the previous paragraph, the P-flatband in our optical lattice is not perfectly flat and is characterised by a small dispersion (curvature) around the $M$ point which is only $\approx5$ $\upmu$eV different in energy from both the $\Gamma$ or the $X$ points (see Fig.~\ref{fig4}i). This small band curvature can trigger condensation into the band minima and is attributed to the finite potential depth of the lattice sites (taken here to be 2 meV in calculation) which leads to deviation from the perfectly dispersionless bands predicted by the tight binding theorem~\cite{Shen_PRB2010}. The band curvature can be reduced by increasing the potential depth of the inverse Lieb lattice (see Sec.~S5 in SM) which can be achieved by adjusting the system properties (e.g., exciton-photon detuning or the exciton dipole moment by appropriate choice of the semiconductor material) such that stronger pumping (bigger blueshifting reservoir) is required to achieve condensation. Another feature of our inverse optical Lieb lattice is that it creates slightly elliptical confinement at sublattices A and C which splits the energies and linewidths of the $P_{x,y}$ orbitals. In other words, the overlap of $P_{x,y}$ orbitals into the pump (gain) region is different. For sublattice A (C) the $P_x$ $(P_y)$ orbital overlaps more with the pump which creates a higher gain for the P-flatband state. This interpretation is supported through a non-Hermitian tight-binding theory (see Sec.~S6 in SM).

\section*{Discussion}
The demonstrated optically arranged system of exciton-polaritons presents a flexible platform to study fundamental proposals on non-Hermitian phenomena in artificial lattices including nonlinear reshaping of the crystal bands when above condensation threshold. In Fig.~\ref{fig2} we have demonstrated the condensation of polaritons into excited Bloch states belonging to a lattice of scatterers (narrow waist repulsive potentials) shaped into a Lieb pattern. The observed crystal bands and agreement with theoretical models opens up a path to explore more intriguing effects of scattered matter-waves such as slow polaritons~\cite{Baba_NatPho2008}, guided polaritons~\cite{Mekis_PRL1996}, and solitonic modes~\cite{Redondo_NatComm2014}. In Fig.~\ref{fig3} we have revealed a gradual nonequilibrium phase transition from the scatterer Lieb lattice, characterised by strongly gain guided and energetic polariton condensates, to a square lattice of optically trapped condensates as a function of two easily tunable parameters in experiment, the lattice constant and pump power. The transition regime is accompanied by multimode polariton lasing which is a unique feature of polariton systems due to their strong interactions, that leads to unexpected condensation patterns in real space. Such multimodal behaviour of the condensate implies close gain competition between gain guided and trapped polaritons. Our observations therefore highlight the intriguing duality of polaritons as ``lasers'' (gain guiding) and ``Bose-Einstein condensates'' (thermalisation) with interesting perspectives as strongly nonlinear objects in non-Hermitian optical lattices. As an example, the optical malleability of our matter-wave platform allows one to study in detail the gradual departure from a system of quantised energies to that of smoothly connected quasimomentum states by building the lattice up cell-by-cell.

Another possible perspective is to optically engineer the lattice to probe exotic band properties for polariton condensation. In Fig.~\ref{fig4} we have demonstrated a conventional (i.e., tightly bound waves) Lieb lattice, by packing pump spots into its inverse shape, with subsequent condensation into flat band states. Given the nonlinear nature of exciton-polaritons, condensation into adjustable flat band states could open a window to investigate strongly correlated states of matter and the effects of disorder against interactions~\cite{Baboux_PRL2016}. The implemented energy tomography methods provide complete access to the polariton states in Fourier space and can be used to study reshaping of the spectrum under arbitrary lattice deformation and nonlinear effects. We point out that the current study is performed in the scalar polariton regime, but can be easily extended to include its spin degree of freedom by changing the pump polarisation. Furthermore, bands induced by our optically engineered landscape can be populated using resonant excitation permitting study of the evolution of polariton matter waves in non-Hermitian optical lattices with any chosen crystal momentum and frequency. We believe that our work carries significant weight in the future design and investigation of polaritonic non-Hermitian (gain and loss) lattice physics in e.g. topological lasers~\cite{Amelio_PRX2020, Comaron_PRR2020}, phase transitions in many body systems~\cite{Zamora_PRX2017, Fruchart_Nature2021}, non-reciprocal transport~\cite{Mandal_PRL2020}, and access to a multitude of gain-induced anomalies reported for diffractive metasurfaces~\cite{Kolkowski_IEEE2020}.

\bibliographystyle{naturemag}

\section*{Acknowledgements}
We acknowledge the support of the UK’s Engineering and Physical Sciences Research Council (grant EP/M025330/1 on Hybrid Polaritonics), the RFBR projects No. 20-52-12026 (jointly with DFG) and No. 20-02-00919. H.S. and P.G.L acknowledge support by the European Union’s Horizon 2020 program, through a FET Open research and innovation action under the grant agreement No. 899141 (PoLLoC).

\section*{Author contributions}
P.G.L. led the research project. P.G.L., S.A., A.A., J.D.T. designed the experiment. S.A., A.A. and J.D.T. carried out the experiment and analysed the data. H.S. developed the theoretical modelling and performed numerical simulations. All authors contributed to the writing of the manuscript.

\section*{Competing interests}
The authors declare no competing interests.

\setcounter{equation}{0}
\setcounter{figure}{0}
\setcounter{table}{0}
\renewcommand{\theequation}{S\arabic{equation}}
\renewcommand{\thefigure}{S\arabic{figure}}
\renewcommand{\thesection}{S\arabic{section}}
\onecolumngrid
\newpage
\vspace{1cm}
\begin{center}
\Large \textbf{Supplementary Information}
\end{center}

\renewcommand{\thefootnote}{\alph{footnote}}	
\addto\captionsenglish{\renewcommand{\figurename}{Supplementary Figure}}

\section*{Supplementary Note 1: Additional experimental measurements on optical lattices with square, Lieb, and inverse Lieb geometry}
In this section we provide additional data on the measured polariton photoluminescence (PL) in the diffractive scatterer regime using pump spots arranged in a square lattice with lattice constant $D =22.5\ \upmu$m (see Supplementary Fig.~\ref{fig.square}) and Lieb lattice $D =22.3\ \upmu$m (see Supplementary Fig.~\ref{fig.biglieb}). We also present complementary data on the inverse Lieb lattice in Fig.~4 in the main manuscript (see Supplementary Fig.~\ref{figpfband}). The full-width-half-maximum (FWHM) of the Gaussian laser spots exciting the condensates is $ \approx 2$ $\upmu$m.  Just like in the main text, we define the real space horizontal and vertical coordinates as $\mathbf{r} = (x,y)$, and in momentum space $\mathbf{k} = (k_x,k_y)$.

Supplementary Figs.~\ref{fig.square}a and~\ref{fig.square}b show the PL in real space and energy-momentum space (along $k_x=0$) approximately at threshold ($P \approx P_\text{thr}$) for a square arranged lattice of widely spaced Gaussian pump spots. At this power no significant phase locking occurs between the condensates as can be evidenced from the absence of interference fringes in the spatial PL and the broad energy distribution of polaritons in the lattice dispersion. In Supplementary Figs.~\ref{fig.square}c and ~\ref{fig.square}d we show the same measurements above condensation threshold ($P \approx 1.3P_\text{thr}$) which result now in clear appearance of interference fringes between the radiating condensates and a greatly narrowed energy distribution peaking around 1.4478 eV. We note that Supplementary Figs.~\ref{fig.square}b and~\ref{fig.square}d are plotted on a logarithmic scale. We observe multiple illuminated high energy bands in the polariton dispersion verifying that polaritons in the system are experiencing crystal scattering in the optical lattice. 
\begin{figure}
	\includegraphics[width = 0.6\linewidth]{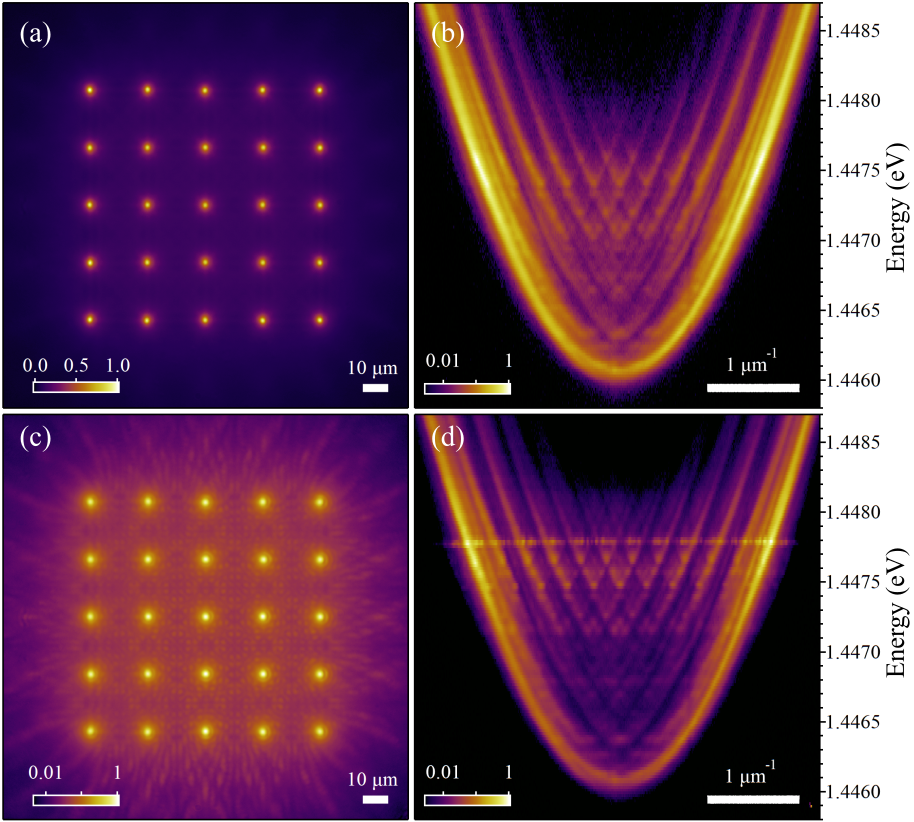}
	\caption{\textbf{Optical lattice of 25 polariton condensates arranged in a square geometry} with a lattice constant set to  $D = 22.5$ $\upmu$m. (a,c) Real space and (b,d) energy-resolved momentum space polariton PL at ($P \approx P_\text{thr}$) and above ($P \approx 1.3 P_\text{thr}$) threshold respectively.}
	\label{fig.square}
\end{figure}

In Supplementary Fig.~\ref{fig.biglieb} we provide similar measurements to those in Fig.~2a-e in the main manuscript but using a larger lattice constant of $D =22.3\ \upmu$m. The results presented in Supplementary Fig.~\ref{fig.biglieb} show that by changing the lattice constant in the scatterer regime we can change the qualitative features of the polariton PL above threshold. Here we observe destructive interference inside the cells of the lattice (i.e., where pump spots are absent) whereas in Fig.~2 (main manuscript) we saw constructive interference. This result can be intuitively understood from the fact that polaritons are condensing into a favorable Bloch mode of the optical lattice by maximising their mutual interference (overlap) over the pump gain region. Changing the lattice constant $D$ naturally affect the interference (i.e., more wavelengths fit between pump spots) and one observes distance-periodic changes in the polariton PL pattern as different Bloch modes move in and out of the lattice gain bandwidth.
\begin{figure}[t]
	\includegraphics[width = 0.95\linewidth]{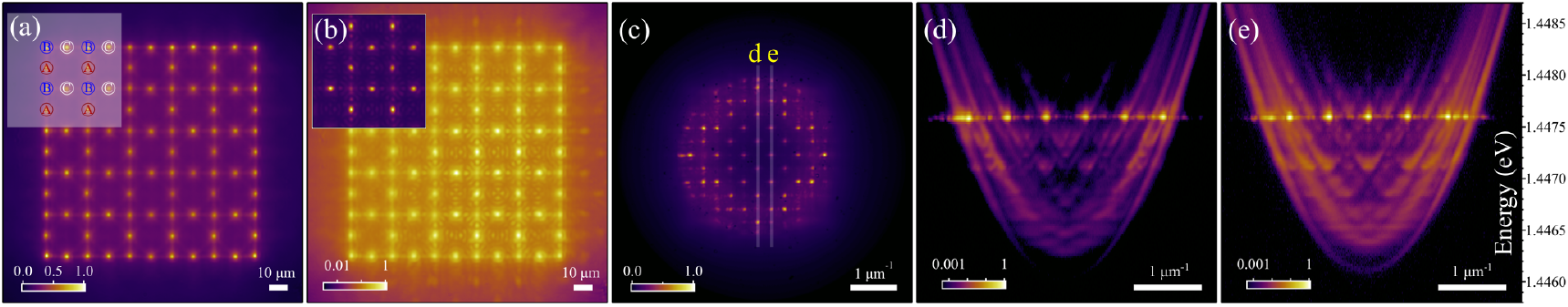}
	\caption{\textbf{Optical lattice of 96 polariton condensates arranged in a Lieb geometry} (a) at threshold with a lattice constant set to  $D = 22.3$ $\upmu$m. The inset in (a) schematically shows sublattices denoted with letters A,B,C forming the Lieb lattice. (b) Real space polariton PL for the lattice pumped at $P \approx 1.1 P_\text{thr}$. Driving the system above threshold leads to the condensation on A and C sublattices, see inset shown in linear colorscale. (c) Momentum space polariton PL for $P \approx 1.1P_\text{thr}$. Semi-transparent thin lines denoted with yellow letters ``d'' and ``e'' correspond to $k_x = 0$ and $k_x = 2\pi/D$, for which the energy-resolved momentum space PL shown in (d),(e) respectively.}
	\label{fig.biglieb}
\end{figure}

In Supplementary Fig.~\ref{figpfband} we provide experimental results for the inverse (conventional tight binding) Lieb lattice hosting $P$-flatbands corresponding to Figs.~4e-g in the main text. Supplementary Fig.~\ref{figpfband}a shows the energy-resolved reciprocal space PL taken at $k_x = 0$ where yellow arrows indicate the location of $P$- and $S$-bands. Supplementary Fig.~\ref{figpfband}b is measured at $k_x = 2\pi/D$ where the two horizontally dashed lines, denoted with yellow letters ``(c)'' and ``(d)'' correspond to the momentum space isoenergy planes PL shown in Supplementary Figs.~\ref{figpfband}c and~\ref{figpfband}d. We note that in each panel the PL is independently normalised. The data shown in Supplementary Fig.~\ref{figpfband}c is the same as shown in Fig.~4l in the main manuscript and corresponds to polariton condensation into the $M$ symmetry point of the $P$-flatband. Unfortunately, the limited energy resolution of our experimental setup and the small non-stimulated population of polaritons in surrounding band structures prevents us from accurately resolving the $P$-flatbands in energy-momentum space. We additionally present the momentum space PL corresponding to the lowest $P$-band, forming an peculiar square shape pattern along the $\Gamma \leftrightarrow X$ edges in the outer Brillouin zones. The lowest $P$-band is completely flat along the $\Gamma \leftrightarrow X$ edges but possesses strong dispersion along the $X \leftrightarrow M \leftrightarrow \Gamma$ edges as shown in Fig.~4i in the main manuscript. The same type of PL is also observed in our calculations (see Supplementary Fig.~\ref{figs3}e).

\begin{figure}[h]
	\includegraphics[width = 0.9\linewidth]{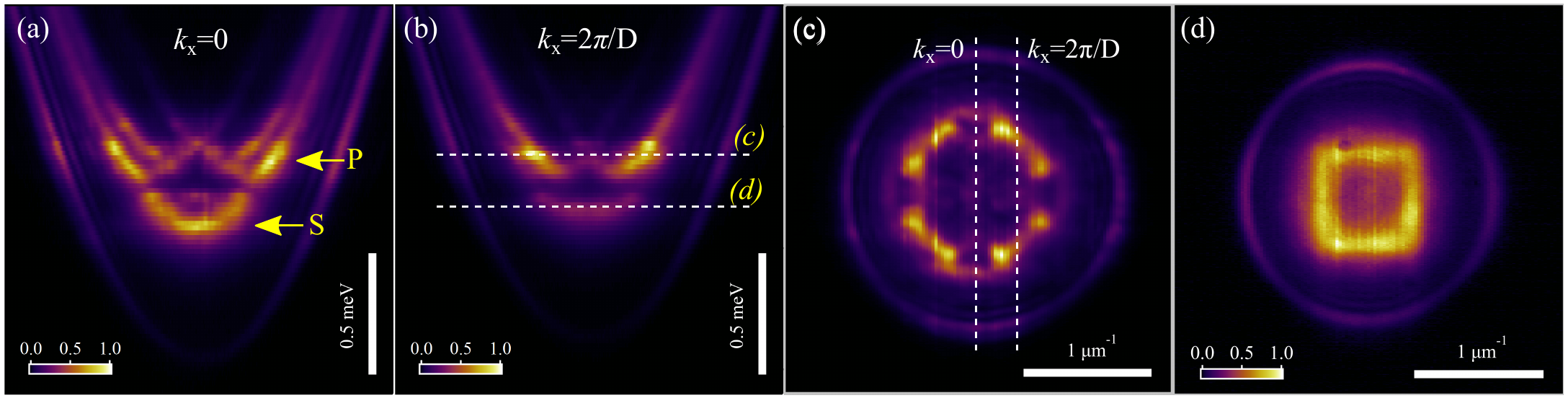}
	\caption{\textbf{Experimentally measured band structure in an inverse Lieb lattice with $D = 16$ $\upmu$m.} (a,b) Energy-resolved momentum space polariton PL at $k_x = 0$ and $k_x = 2\pi/D$ corresponding to the system shown in Figs.~4e-g of the main text. Yellow arrows in (a) indicate $P$-band and $S$-band states respectively. White dashed lines in (b) with yellow letters ``(c)'' and ``(d)'' specify the energies of the momentum space PL presented in (c) and (d), respectively. All panels are normalised independently.}
	\label{figpfband}
\end{figure}

\section*{Supplementary Note 2: Theory of polariton dynamics} \label{sec1}
We consider the two-dimensional (2D) Schrödinger equation which describes a polariton in the static potential landscape designed by our optical non-resonant lasers.
\begin{equation} \label{eq.schro}
i  \hbar \frac{d \Psi}{d t}  = \left[ \hat{E} + V(\mathbf{r}) \right] \Psi
\end{equation}
Here, $\hat{E}$ is the dispersion of the lower polariton branch. The upper polariton branch is neglected because it is both far away in energy and weakly populated by our experimental excitation scheme. Given the much larger effective mass of the excitons compared to the cavity photons we can set their dispersion to zero and the lower polariton dispersion in momentum space becomes~\cite{SM_kavokin2017microcavities},
\begin{equation} \label{eq.disp}
\hat{E}(\mathbf{k}) = \frac{1}{2} \left[\hat{E}_\phi(\mathbf{k}) - \sqrt{ \hat{E}_\phi^2(\mathbf{k}) + (\hbar \Omega)^2 } \right],
\end{equation}
where $\hat{E}_\phi(\mathbf{k})$ is the cavity photon dispersion and $\hbar \Omega = 12$ meV is the Rabi energy of our sample~\cite{SM_cilibrizzi_polariton_2014}. The longitudinally confined photons obey the dispersion relation,
\begin{equation}
\hat{E}_\phi(\mathbf{k}) = \sqrt{ \left(\frac{hc}{\lambda}\right)^2 + \left( \frac{\hbar c k}{n}\right)^2} + \Delta - \frac{i \hbar \gamma}{2},
\end{equation}
where $\lambda = 860$ nm is the cavity resonance, $n = 3.556$ is the cavity refractive index,  $\Delta = -5$ meV is the cavity photon detuning from the exciton resonance, and $\gamma = 1/5.5$ ps$^{-1}$ is the cavity photon lifetime. The real part of the calculated dispersion is shown with a white dashed line plotted against the experimentally retrieved dispersion in Supplementary Fig.~\ref{scheme}a. In Supplementary Fig.~\ref{scheme}b we show a schematic of the Lieb lattice and the three square sublattices denoted by their label $A,B,C$.
\begin{figure}[b]
	\includegraphics[width = 0.75\linewidth]{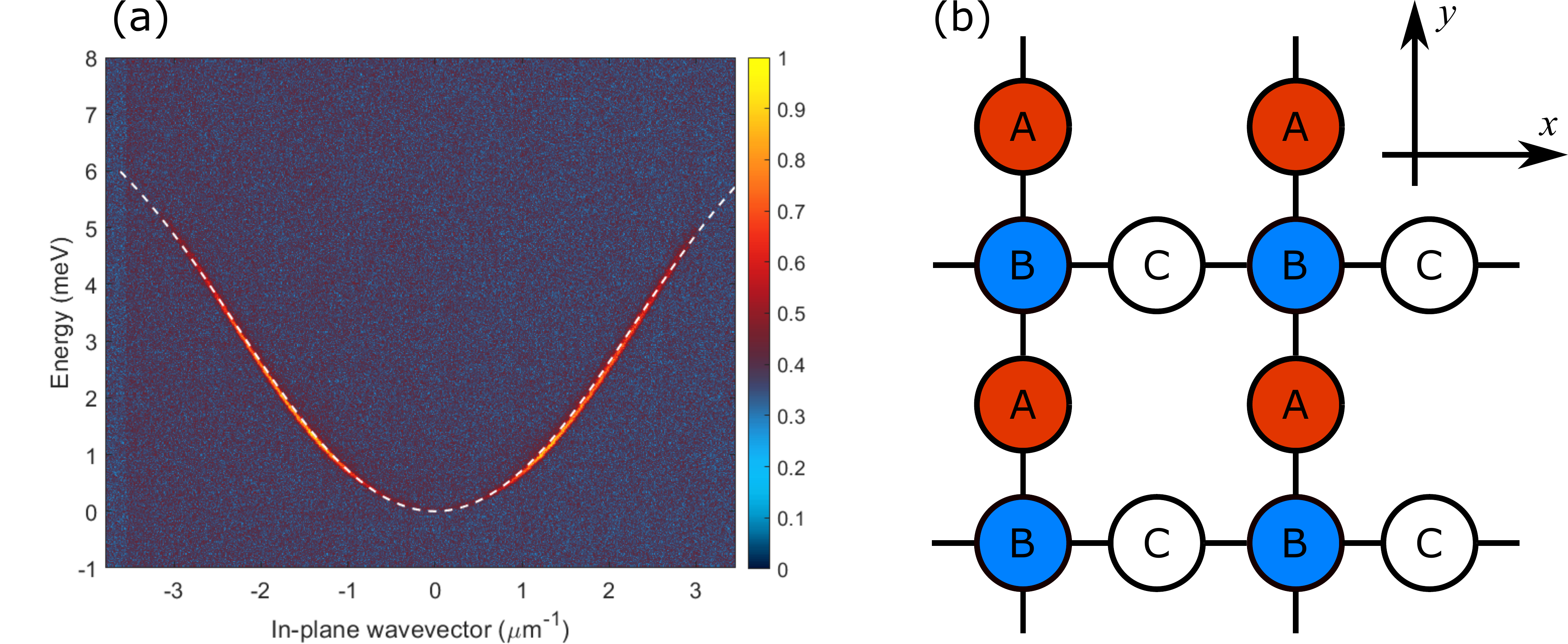}
	\caption{\textbf{Bare lower polariton dispersion and Lieb lattice illustration.} (a) Heatmap shows the normalized measured lower polariton dispersion with Eq.~\eqref{eq.disp} plotted on top as a white dashed line. (b) Schematic of the Lieb lattice geometry consisting of three square sublattices labeled $A,B,C$. }
	\label{scheme}
\end{figure}

The potential $V(\mathbf{r})$ is directly proportional to the incident non-resonant laser intensity and can be written as a superposition of spatially separated Gaussians,
\begin{equation} \label{eq.pump}
V(\mathbf{r}) = V_0 \sum_{n=1}^N \exp{\left[-\frac{(x - x_n)^2 + (y-y_n)^2}{2w^2}\right]}.
\end{equation}
Here, $N$ is the number of pumps in question and $x_n, y_n$ are their coordinates, $V_0$ denotes the complex potential strength. $\text{Re}{(V_0)}>0$ and $\text{Im}{(V_0)}>0$ denote the blueshift and particle gain experienced by polaritons respectively due to their interactions with the laser generated background of non-condensed particles, and $w = 1.274$ $\upmu$m is the RMS width of the pump spots corresponding to 3 $\upmu$m FWHM. The width is slightly larger than that of the lasers (2 $\upmu$m FWHM) to account for the finite diffusion of excitons in the reservoir from the laser spot. 

The dispersion of Eq.~\eqref{eq.schro} can be numerically obtained through Monte-Carlo methods where an average is performed over the time evolution of many random initial realizations of the particle wavefunction $\Psi(\mathbf{r},t=0)$. This method is particularly useful for potentials which lack the usual symmetries required by Bloch's theorem to work. It permits calculation over which frequency and momentum components $\tilde{\psi}(\omega,\mathbf{k})$ survive the longest in the system (i.e., decay the slowest) in presence of optical gain at the pump spots [i.e., $\text{Im}{(V_0)}>0$]. Results of this method are  shown in Supplementary Figs.~\ref{figs1}-\ref{figs3} where we extract numerically the dispersion from the dynamics of many random initial realizations of $\tilde{\psi}{(\omega,\mathbf{k})}$ in Eq.~\eqref{eq.schro} in several lattice geometries relevant to the study. Overall we find good agreement between experimental observations and our calculations across all lattice geometries.
\begin{figure}[b]
	\includegraphics[width = \linewidth]{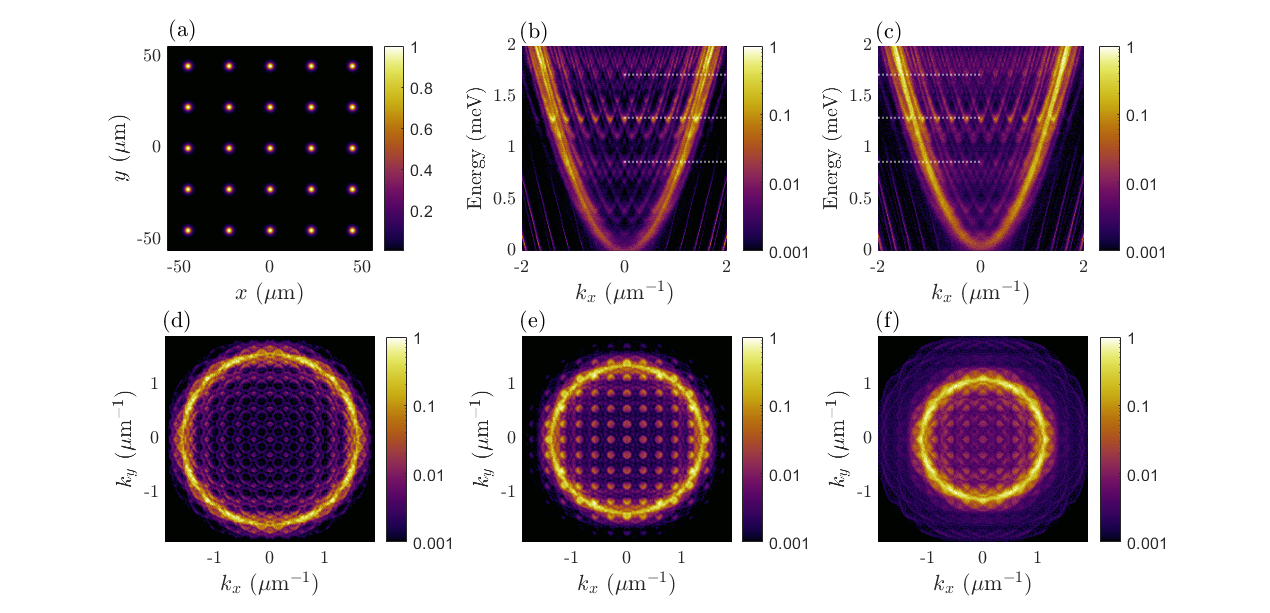}
	\caption{\textbf{Calculated band structure in a square arranged scatterer lattice of Gaussian spots.} (a) Non-resonant laser profile $V(\mathbf{r})$ (normalised) shaped into a square scatterer lattice of Gaussians with a lattice constant $D = 22.5$ $\upmu$m and $V_0 = 2 + 0.25i$ meV. (b,c) Energy-momentum space PL at (b) $k_y = 0$ and (c) $k_y = 2\pi/D$. (d-f) Polariton momentum space PL around threshold at $E = 1.72, 1.3, 0.87$ meV respectively corresponding to the white dotted lines in (b,c).}
	\label{figs1}
\end{figure}
\begin{figure}
	\includegraphics[width = \linewidth]{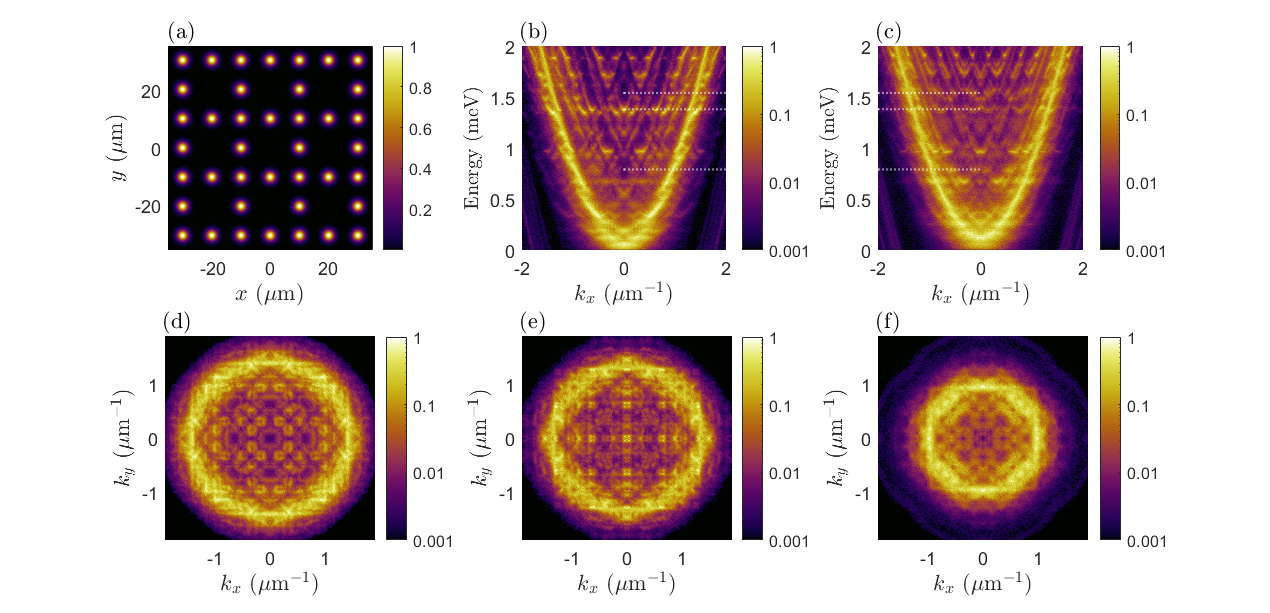}
	\caption{\textbf{Calculated band structure in a Lieb arranged scatterer lattice of Gaussian spots.} (a) Non-resonant laser profile $V(\mathbf{r})$ (normalised) shaped into a Lieb scatterer lattice of Gaussians with a lattice constant $D = 20.3$ $\upmu$m and $V_0 = 2.2 + 0.25i$ meV corresponding to Fig.~2 in the main text. (b,c) Energy-momentum space PL around threshold for (b) $k_y = 0$ and (c) $k_y = 2\pi/D$. (d-f) Polariton momentum space PL around threshold at $E = 1.54, 1.38, 0.79$ meV respectively corresponding to the dotted lines in (b,c).}
	\label{figs2}
\end{figure}
\begin{figure}
	\includegraphics[width = \linewidth]{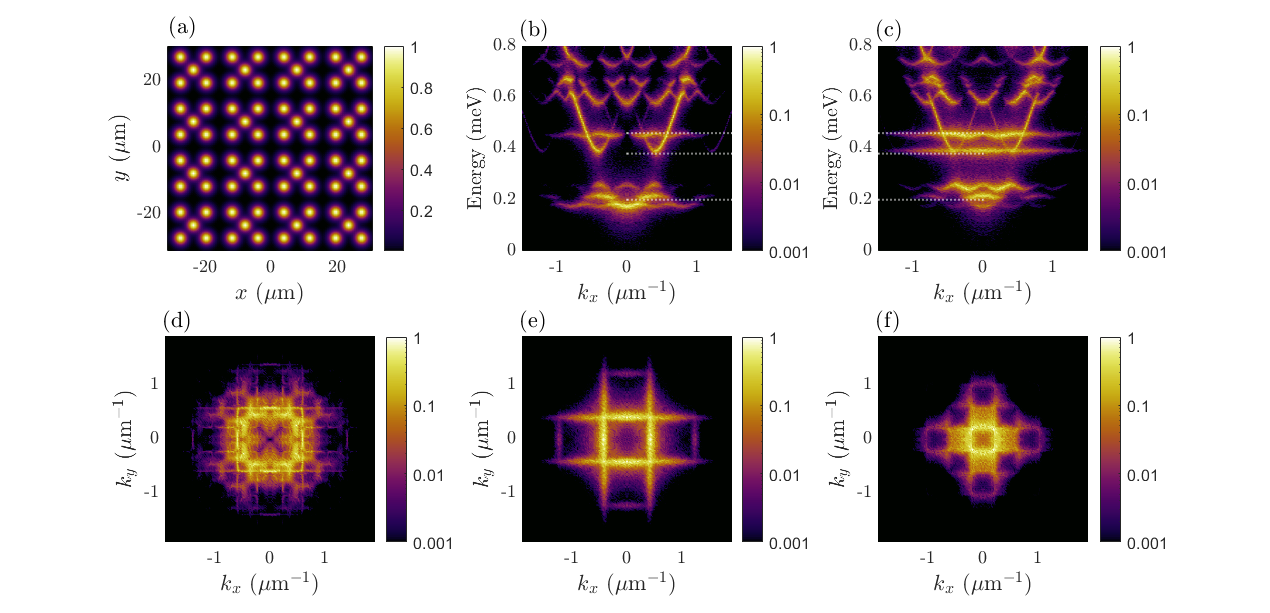}
	\caption{\textbf{Calculated band structure in a Lieb lattice formed from inverse arrangement of Gaussian pump spots.} (a) Non-resonant laser profile shaped to form a conventional Lieb lattice of potential minima between the lasers. Lattice constant is $D = 15.6$ $\upmu$m and $V_0 = 2$ meV corresponding to Fig.~4 in the main text. (b,c) Energy-momentum space PL around threshold for (b) $k_y = 0$ and (c) $k_y = 2\pi/D$. (d-f) Polariton momentum space PL around threshold at $E = 0.47, 0.38, 0.2$ meV respectively corresponding to the dotted lines in (b,c).}
	\label{figs3}
\end{figure}

\section*{Supplementary Note 3: Bloch theorem} \label{sec.bloch}
\begin{figure}[t!]
\centering
\includegraphics[width=\columnwidth]{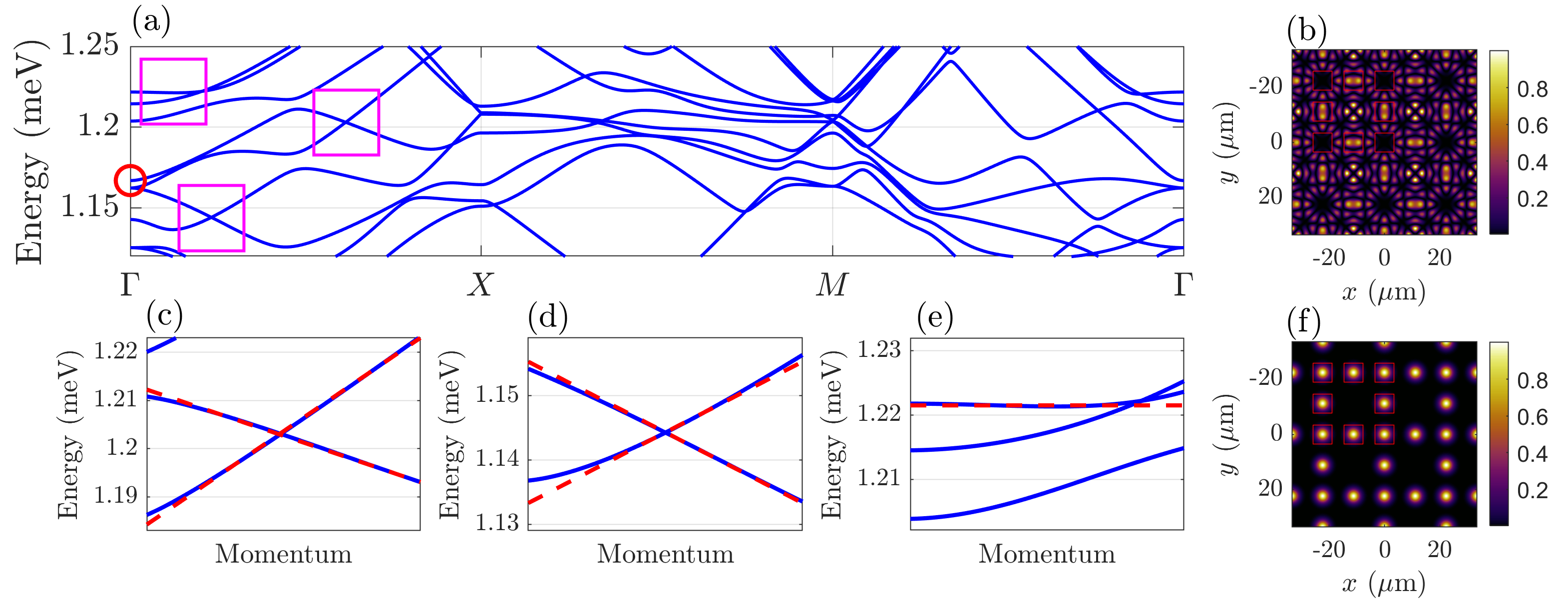}
\caption{\textbf{Calculated polariton bands using Bloch's theorem in the Lieb scatterer lattice.} (a) Calculated bands using Eq.~\eqref{eq.bloch_eig} for a Lieb lattice of Gaussians, similar to Fig.~\ref{figs2} and Fig.~\ref{fig.biglieb}. (b) Calculated Bloch state intensity $|u_{n_c,\mathbf{q}=0}(\mathbf{r})|^2$ belonging to the band marked by the red circle in (a) and the condensate real space PL from Fig.~2(b) in the main text. The arrangement of the Gaussian pumps is indicated by the red squares. (c-e) Zoomed in regions of the dispersion showing exotic points where the dispersion becomes practically linear (massless polaritons) or flat (heavy polaritons). (f) Potential profile $V(\mathbf{r})$ (normalised) shaped into a Lieb scatterer lattice of Gaussians with a lattice constant $D = 22.3$ $\upmu$m and $V_0 = 1.25$ meV.}
\label{bloch1}
\end{figure}

In this section we investigate the dispersion belonging to Eq.~\eqref{eq.schro} by applying Bloch's theorem. The following equations were used to calculate the bands plotted in Figs.~2i and~4i in the main text. For simplicity, we apply a parabolic approximation by writing,
\begin{equation}
\hat{E}(\mathbf{k}) = \frac{\hbar^2 k^2}{2m}
\end{equation}
where the effective mass of the lower polaritons is $m = 0.32$ meV ps$^2$ $\upmu$m$^{-2}$ found by fitting the dispersion in Supplementary Fig.~\ref{scheme}a. The potential is assumed to be infinite and periodic such that $V(\mathbf{r}) = V(\mathbf{r}+\mathbf{D})$ where $\mathbf{D} = n_1 \mathbf{a}_1 + n_2 \mathbf{a}_2$ is the translational symmetry vector defined in the bases of primitive lattice vectors $\mathbf{a}_{1,2}$ for some integers $n_{1,2}$. We can then apply Bloch's theorem where we write the polariton single particle wavefunction in the factorised form of crystal momentum $\mathbf{q} = (q_x,q_y)$ and Bloch states in the $n$th band $u_{n,\mathbf{q}}(\mathbf{r})$,
\begin{equation} \label{eq.bloch}
\Psi_n(\mathbf{r}) = e^{i \mathbf{q} \cdot \mathbf{r}} u_{n,\mathbf{q}}(\mathbf{r}).
\end{equation}
Here, $u_{n,\mathbf{q}}(\mathbf{r}) = u_{n,\mathbf{q}}(\mathbf{r}+\mathbf{D})$. Substituting Eq.~\eqref{eq.bloch} into Eq.~\eqref{eq.schro} the Bloch eigenvalue problem then reads,
\begin{equation}\label{eq.bloch_eig}
\left[ \frac{\hbar^2}{2m}\left[ \left(q_x - i \frac{\partial}{\partial x}\right)^2 + \left(q_y - i \frac{\partial}{\partial y}\right)^2 \right] + V(\mathbf{r}) \right] u_{n,\mathbf{q}}(\mathbf{r}) = E_{n,\mathbf{q}} u_{n,\mathbf{q}}(\mathbf{r}).
\end{equation}
We can diagonalise Eq.~\eqref{eq.bloch_eig} by expressing the Gaussian potential as a Fourier series in the basis of the reciprocal lattice vectors defined through $\mathbf{G} \cdot \mathbf{D} =2\pi n$ for $n \in \mathbb{Z}$,
\begin{equation}
V(\mathbf{r}) = \sum_{G} V_{G} e^{i \mathbf{G} \cdot \mathbf{r}}.
\end{equation}
Supplementary Fig.~\ref{bloch1}a shows a portion of the calculated band structure for a Lieb arrangement of the Gaussian spots (see Supplementary Fig.~\ref{bloch1}f) with a lattice constant of $D = 22.3$ $\upmu$m with some features of interest marked with magenta colored boxes. Supplementary Fig.~\ref{bloch1}b shows the real space intensity of the Bloch state $|u_{n_c,\mathbf{q}=0}|^2$, marked by red circle corresponding to the experimentally observed pattern shown in Supplementary Fig.~\ref{fig.biglieb}. A portion of the Gaussian pump locations are marked by red squares in both Supplementary Figs.~\ref{bloch1}b and~\ref{bloch1}f for clarity.  In Supplementary Fig.~\ref{bloch1}a we observe regions in the band structure with almost linear- and flat-like bands (magenta squares zoomed in Supplementary Figs.~\ref{bloch1}c-e). Although such exotic points were not experimentally investigated in this study, their theoretical prediction enhances the versatility and importance of the investigated scatterer Lieb lattice. Even exciting polaritons into such points via non-resonant methods is infeasible they can still be excited using resonant lasers. This can potentially open up investigation into nonlinear transport properties of nearly massless or slow-light polaritons. 

We also observe that the calculated Bloch state shown in Supplementary Fig.~\ref{bloch1}b shows complete destructive interference at sublattice $B$ whereas in the experimentally measured spatial PL in Supplementary Fig.~\ref{fig.biglieb}b this destructive interference is not complete. This could mean the the condensate is populating more than one Bloch state of the lattice since Supplementary Fig.~\ref{bloch1}a clearly shows that multiple bands exist within a narrow energy range (i.e., many bands can be within the gain bandwidth of the lattice). The time-integrated measurements of our experiment would then instead show a weighted average over these multiple Bloch states which explains the discrepancy between linear Bloch's theory and our measurements. In order to get better agreement between measurements and theory, we introduce in the next section a generalised Gross-Pitaevskii model which describes the polarion dynamics above condensation threshold (i.e., in the nonlinear regime). 

\section*{Supplementary Note 4: Generalised Gross-Pitaevskii simulations}
We simulate the dynamics of the polariton condensate in the parabolic regime of the lower polariton branch using the two-dimensional driven-dissipative (i.e., generalised) Gross-Pitaevskii model~\cite{SM_Wouters_PRL2007} written for the condensate wavefunction, $\Psi({\bf r},t),$ and a rate equation for the exciton reservoir supplying particles into the condensate, $n({\bf r},t)$:
\begin{align} \label{Eq.GPE}
i \frac{\partial \Psi}{\partial t} &= \Big[ \frac{-\hbar \nabla^2}{2m} + \alpha |\Psi|^2 -\frac{i\gamma}{2} + \left(g+\frac{iR}{2}\right)n + G P(\mathbf{r})\Big] \Psi + i  \mathfrak{R}(\Psi), \\ \label{eq.res}
\frac{\partial n}{\partial t} &= -(\Gamma_R + R|\Psi|^2)n + P(\mathbf{r}).
\end{align} 
Here, $m$ is the polariton effective mass, $\alpha$ is the polariton-polariton interaction strength, $R$ is the scattering rate of reservoir excitons into the condensate, $\gamma$ is the rate of polariton losses through the cavity mirrors, $\Gamma_R$ is the decay rate of the reservoir excitons, $g$ and $G$ are the interaction strengths of polaritons with the exciton reservoir feeding the condensate (i.e., so-called {\it bottleneck} or {\it active} excitons), and high momentum photoexcited excitons $P(\mathbf{r})$ (proportional to the laser profile and power) that do not scatter into the condensate, respectively. The final term $\mathfrak{R}(\Psi)$ is an energy relaxation term~\cite{SM_Wouters_PRB2010, SM_PhysRevB.92.035305} that assists ground state condensation and is taken proportional to the background density of bottleneck excitons,
\begin{equation}
    \mathfrak{R}(\Psi) =  (\lambda n)^\xi \left[ \frac{\hbar \nabla^2}{2m}+ \mu(\mathbf{r},t) \right] \Psi.
\end{equation}
Here, $\lambda$ denotes the energy relaxation efficiency, $\xi$ is a fitting parameter, and $\mu(\mathbf{r},t)$ is the particle conserving local effective chemical potential of the condensate.

To take into account to finite diffusion of the reservoir excitons $n(\mathbf{r},t)$ away from the laser pump spots we approximate the reservoir driving profile with a Lorentzian,
\begin{equation} 
P(\mathbf{r}) = P_0 \sum_{n=1}^N \frac{L^2}{L^2 + (x - x_n)^2 + (y-y_n)^2}.
\end{equation}
Here, $N$ is the number of pumps in question and $x_n, y_n$ are their coordinates, $2L = 2$ $\mu$m denotes the full width at half maximum of the reservoir driving field and the value $P_0$ denotes the laser power density. We point out that the band features previously analysed in Sec.~\ref{sec1} using lattices made up of arrranged Gaussian spots are not qualitatively changed here. While our model aims at being as simple as possible to produce the observed effects, more accurate simulations could possibly be achieved by introducing explicitly an exciton diffusion term into Eq.~\eqref{eq.res}~\cite{SM_Tosi_NatPhys2012} and stochastic noise~\cite{SM_toepfer_coherence2020}. The lower polariton mass and lifetime are taken corresponding to the properties of our cavity: $m = 0.32$ meV ps$^2$ $\mu$m$^{-2}$, and $\gamma = 1/5.5$ ps$^{-1}$. The remaining parameters are taken similar to those used in previous works: $\hbar\alpha = 3.3$ $\mu$eV $\mu$m$^{2}$, $g = \alpha$, $G = 2 g/\Gamma_R$, $R = 0.8 \alpha$, $\Gamma_R = 2 \gamma$, $\xi = 1/4$, and $\lambda^\xi = 0.004$ $\mu$m$^2$.

In each simulation we apply stochastic initial conditions to $\Psi(\mathbf{r},t=0)$ and run the simulation much longer than the characteristic timescales of the condensate dynamics. This typically corresponds to simulation times longer than $10$ ns. The results of such simulations are presented in the insets of Figs.~3a-c and~3e in the main text where we have plotted the time-integrated (averaged) condensate density $\langle |\Psi(\mathbf{r},t)|^2 \rangle$ at the same lattice constants as in experiment for pump powers $P_0$ chosen to give the best match. The single exception is the data presented in the inset in Fig.~3d which was not obtained from Gross-Pitaevskii simulations. This was due to the lack of the simulated condensate stabilising in a state similar to the one observed in experiment for the chosen parameters. It is important to stress that this does not preclude such a condensed state existing in stable form in simulation for other parameters. However, we are able to find a good match using Bloch's theorem [see Eq.~\eqref{eq.bloch_eig}] where we plot in Fig.~3d the state $|\Psi(\mathbf{r})|^2 = |u_{3,\mathbf{q}_X}(\mathbf{r})|^2 + |u_{3,\mathbf{q}_{X'}}(\mathbf{r})|^2$ with $\mathbf{q}_X = (0,\pi/D)$ and $\mathbf{q}_{X'} = (\pi/D,0)$ corresponding to the edges of the Brillouin zone. 

In Fig.~3m in the main manuscript we qualitatively produce the experimental observation of gain guided condensates collapsing into trapped condensates at high power and small lattice constant through Gross-Pitaevskii simulations. Some discrepancy can be observed between simulation and experiment at low powers and small lattice constants where the simulation overestimates the gain guided polaritons. This discrepancy can appear from the lack of stochastic treatment (i.e., Wiener noise) in our equations which would smear out the simulated condensate PL at low powers close to threshold.

\section*{Supplementary Note 5: Finite $P$-flatband curvature}
As shown in Fig.~4i and~4j in the main manuscript, there exist residual finite band curvature in the $P$-flatbands of the inverse optical Lieb lattice. As mentioned in the text of of main manuscript the dispersion (curvature) of the flatband appears around the $M$ point and is $\approx5$ $\upmu$eV different in energy from both the $\Gamma$ or the $X$ points (see Supplementary Fig.~\ref{fig.s7}b). Interestingly, the curvature of one flatband corresponds to lattice particles of negative effective mass and the other to positive effective mass. Intuitively, the nonequilibrium nature and energy relaxation of polaritons triggers condensation into the minima of the lower flatband. The source of this curvature is attributed to the finite potential depth and the Gaussian shape of the excitation beams (see Supplementary Fig.~\ref{fig.s7}a) which leads to deviation from the perfectly dispersionless bands predicted by the tight binding theorem. By increasing the depth of the potential minima (or conversely, the height of the potential maxima) in the lattice the flatband curvature is reduced as can be seen in Supplementary Fig.~\ref{fig.s7}c. The flatbands in Supplementary Fig.~\ref{fig.s7}c are also blueshifted (with respect to the other $P$ bands) which comes from a slightly increased energy splitting between the $P_x$ and $P_y$ orbitals at sublattices A and C due to the Gaussian construction of our lattice.
\begin{figure}
\centering
\includegraphics[width=\columnwidth]{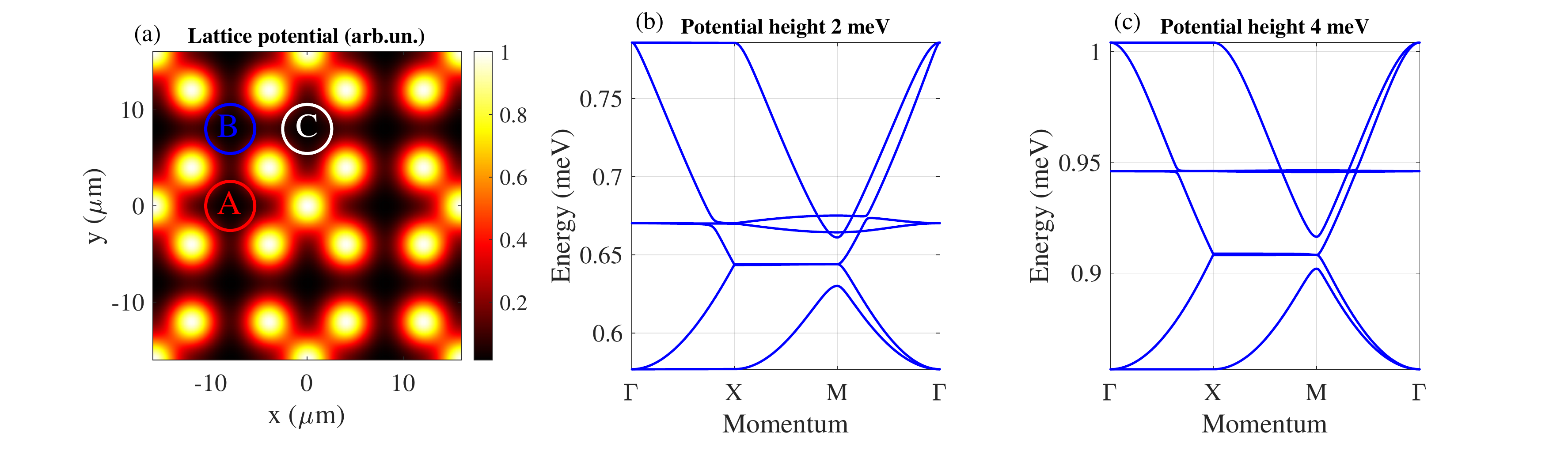}
\caption{\textbf{Effects of the inverse optical Lieb lattice potential height on the P-bands.} (a) Four unit cells of the Pump induced potential landscape used in Bloch’s analysis. (b) Calculated band structure for peak potential height of 2 meV. (c) Calculated band structure for peak potential height of 4 meV. The polariton mass is set to $m = 0.32$ meV ps$^2$ $\mu$m$^{-2}$.}
\label{fig.s7}
\end{figure}

\section*{Supplementary Note 6: Tight binding treatment} \label{sec.tb}
Here, we show qualitatively why condensation preferably occurs into a flatband P-orbital based by considering a tight-binding treatment of the $P$-orbital states. We consider only nearest neighbor coupling and negligible mixing with neighboring $S$- and $D$-orbitals. Being a scalar problem, we also do not consider the polarization degree of freedom of the polaritons. The bulk Hamiltonian in the basis of single particle, lattice site, creation and annihilation operators can be written,
\begin{align} \notag
\hat{H}  = & \sum_{n,m} \bigg[ (\hat{b}^{(x)}_{n,m})^\dagger [J_2(\hat{a}_{n,m}^{(x)} + \hat{a}_{n-1,m}^{(x)}) - J_1 (\hat{c}_{n,m}^{(x)} + \hat{c}_{n,m-1}^{(x)}) ] + \text{h.c.} \\ \notag
 & +  (\hat{b}^{(y)}_{n,m})^\dagger [J_1(\hat{a}_{n,m}^{(y)} + \hat{a}_{n-1,m}^{(y)}) - J_2 (\hat{c}_{n,m}^{(y)} + \hat{c}_{n,m-1}^{(y)}) ] + \text{h.c.} \\ \notag
& +  \epsilon_{a,x} (\hat{a}_{n,m}^{(x)})^\dagger \hat{a}_{n,m}^{(x)} + \epsilon_{b,x} (\hat{b}_{n,m}^{(x)})^\dagger \hat{b}_{n,m}^{(x)} + \epsilon_{c,x} (\hat{c}_{n,m}^{(x)})^\dagger \hat{c}_{n,m}^{(x)} \\
& +  \epsilon_{a,y}(\hat{a}_{n,m}^{(y)})^\dagger \hat{a}_{n,m}^{(y)} + \epsilon_{b,y}(\hat{b}_{n,m}^{(y)})^\dagger \hat{b}_{n,m}^{(y)} + \epsilon_{c,y}(\hat{c}_{n,m}^{(y)})^\dagger \hat{c}_{n,m}^{(y)} \bigg].
\end{align}
Here, $\hat{a}^{(x,y)}_{n,m}, \hat{b}^{(x,y)}_{n,m}, \hat{c}^{(x,y)}_{n,m}$ denote the annihilation operators at sublattices $A,B,C$ respectively [see Supplementary Fig.~\ref{scheme}], $(n,m)$ denotes the site index of the corresponding sublattice, and $(x,y)$ P-orbital lobe alignment along the two orthogonal coordinates $x,y$ of the lattice. The operators obey bosonic commutation relation $[\hat{a}_{n,m}^{(\alpha)}, (\hat{a}_{n',m'}^{(\beta)})^\dagger] = \delta_{\alpha,\beta} \delta_{n,n'} \delta_{m,m'}$ and $[\hat{a}_{n,m}^{(\alpha)}, \hat{a}_{n',m'}^{(\beta)}] = 0 $. The coefficients $J_{1,2} \in \mathbb{C}$ denote the polariton non-Hermitian hopping amplitudes between P-orbitals with orbital axes parallel ($\sigma$-bond) and perpendicular ($\pi$-bond) to the bond axis respectively. The coefficients $\epsilon_{N,\alpha}$ denote on-site energy of the two distinct P-orbitals at each sublattice. 

Performing the standard Fourier transform of the single particle operators one can transform the problem from the basis of lattice sites ($n,m$) into the basis of crystal momentum $\mathbf{q} = (q_x,q_y)$.
\begin{equation} \label{eq.tbq}
\frac{\hat{H}_\mathbf{q}}{2} = \begin{pmatrix} 
\hat{H}_{P_x} & 0_{3,3} \\
0_{3,3} & \hat{H}_{P_y}
\end{pmatrix},
\end{equation}
where
\begin{align*}
\hat{H}_{P_x} & = \begin{pmatrix} 
\epsilon_{a,x} &  J_2  e^{i \frac{q_y}{2}} \cos{(\frac{q_y}{2})} & 0    \\
J_2 e^{- i \frac{q_y}{2}} \cos{(\frac{q_y}{2})} & \epsilon_{b,x} & J_1e^{i \frac{q_x}{2}} \cos{(\frac{q_x}{2})}  \\
0 &  J_1e^{-i \frac{q_x}{2}} \cos{(\frac{q_x}{2})} & \epsilon_{c,x} 
\end{pmatrix}, \\
\hat{H}_{P_y} & = \begin{pmatrix}
 \epsilon_{a,y} &  J_1  e^{i \frac{q_y}{2}} \cos{(\frac{q_y}{2})} & 0  \\
 J_1 e^{- i \frac{q_y}{2}} \cos{(\frac{q_y}{2})} & \epsilon_{b,y} & J_2e^{i \frac{q_x}{2}} \cos{(\frac{q_x}{2})} \\
 0 &  J_2e^{-i \frac{q_x}{2}} \cos{(\frac{q_x}{2})} & \epsilon_{c,y}.
\end{pmatrix}.
\end{align*}
Setting all diagonal terms to zero corresponds to all sites being in resonance and with equal on-site gain. In Supplementary Fig.~\ref{fig.tb}a we plot the dispersion from Eq.~\eqref{eq.tbq} for $J_1 = 0.4 + i0.04$ and $J_2 = J_1/3$. The blue-red colorscale indicates energies with low and high particle gain (imaginary part of the energy), respectively. In this case it shows that polaritons will survive the longest around the $\Gamma$ point in the upper bands. 
\begin{figure}
\centering
\includegraphics[width=0.6\columnwidth]{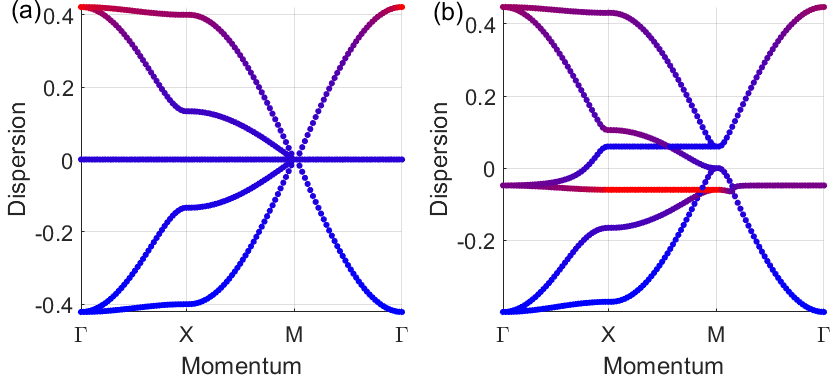}
\caption{\textbf{Calculated P-orbital bands using tight binding theorem.} Blue to red colorscale shows energies of small and large imaginary part (gain) respectively. (a) All diagonal elements $\epsilon_{N,\alpha} = 0$ resulting in highest gain at the top of the conduction $P$-bands. (b) Non-zero diagonal elements introducing splitting between $P_x$ and $P_y$ orbitals at sublattices $A$ and $C$ results in gain moving towards the flatband.}
\label{fig.tb}
\end{figure}

On the other hand, if the $x$ and $y$ $P$-orbitals become split in both real energy and gain at their respective sublattice sites then the dispersion dramatically changes. In Supplementary Fig.~\ref{fig.tb}b we show the calculated energies again for $J_1 = 0.4 + i0.04$ and $J_2 = J_1/3$ but with $x$ and $y$ $P$-orbitals split at sublattices $A$ and $C$. This is modeled by setting $\epsilon_{a,x} = \epsilon_{c,y} = -\epsilon_0 + i \kappa$, $\epsilon_{a,y} = \epsilon_{c,x} = \epsilon_0 - i \kappa$, and $\epsilon_{b,x} = \epsilon_{b,y} = 0$ where $\epsilon_0, \kappa > 0$. Such splitting arises from the slightly elliptically shaped $A$ and $C$ sublattice sites [see Supplementary Fig.~\ref{figs3}a]. Physically, $\kappa$ captures the increased gain of the lower energy state due to greater overlap with the gain region from the Gaussian pumps. In contrast to Supplementary Fig.~\ref{fig.tb}a the flatband between the $X$ and the $M$ points in Supplementary Fig.~\ref{fig.tb}b has now obtained the highest gain due to this non-Hermitian splitting between the $x,y$ $P$-orbitals in the $A$ and $C$ sublattices.

\bibliographystyle{naturemag}

\end{document}